\documentclass{JINST}

\usepackage{amsfonts}
\usepackage{amssymb}
\usepackage{graphicx}
\usepackage{epstopdf}
\usepackage{verbatim}
\usepackage{amsmath}
\usepackage{wasysym}
\usepackage[thickspace,amssymb]{SIunits}

\title{Precise measurement of prompt photon emission for carbon ion therapy}

\author{C.~Agodi$^{f}$, F.~Bellini$^{a,b}$, G.A.P.~Cirrone$^{f}$, F.~Collamati$^{a,b}$, G.~Cuttone$^{f}$, E.~De Lucia$^{c}$, M.~De Napoli$^{f}$, A.~Di Domenico$^{a,b}$, R.~Faccini$^{a,b}$, F.~Ferroni$^{a,b}$, S.~Fiore$^{a,b}$, P.~Gauzzi$^{a,b}$, E.~Iarocci$^{c,d}$, M.~Marafini$^{a,e}$, I.~Mattei$^{a,b}$, A.~Paoloni$^{c}$, V.~Patera$^{c,d}$, L.~Piersanti$^{c,d}$, F.~Romano$^{e,f}$, A.~Sarti$^{c,d}$, A.~Sciubba$^{c,d}$, C.~Voena$^{a,b}$\\
\llap{$^a$} Dipartimento di Fisica, Sapienza Universit\`a di Roma, Roma, Italy \\
\llap{$^b$} INFN Sezione di Roma, Roma, Italy \\
\llap{$^c$} Laboratori Nazionali di Frascati dell'INFN, Frascati, Italy\\ 
\llap{$^d$} Dipartimento di Scienze di Base e Applicate per Ingegneria, Sapienza Universit\`a di Roma,  Roma, Italy\\
\llap{$^e$} Museo Storico della Fisica e Centro Studi e Ricerche ``E.~Fermi'', Roma, Italy\\
\llap{$^f$} Laboratori Nazionali del Sud dell'INFN, Catania, Italy}

\abstract{Proton and carbon ion therapy is an emerging technique used for the treatment of solid cancers. The monitoring of the dose delivered during such treatments is still a matter of research. A possible technique exploits the information provided by single photon emission from nuclear decays induced by the irradiation. This paper reports the measurements of the spectrum and rate of such photons produced from the interaction of a $80\ \mega\electronvolt/$u fully stripped carbon ion beam  at the Laboratori Nazionali del Sud of INFN, Catania, with a  Poly-methyl methacrylate target. 
The differential production rate for photons with energy $E>2\ \mega\electronvolt$ and emitted at  $90^o$ is found to be $dN_{\gamma}/(dN_C d\Omega)=(2.92\pm 0.19)\times 10^{-2}$sr$^{-1}$.
}


\begin{document}

\section*{Introduction}

In the last decade, the use of proton and carbon beams has become more and more widespread as an effective therapy for the treatment of solid cancer (hadrontherapy). Due to their very favorable profile of the release dose in tissue, the hadron beams can be very effective in destroying the tumor and sparing the adjacent healthy tissue in comparison to the standard X-ray based treatment~\cite{Amaldi}. On the other hand, the space selectivity of the hadrontherapy asks for a new approach to the delivered dose monitoring.\\
The uncertainty on the position of the dose released in hadrontherapy treatment can be due to various factors: calibration of the Computed Tomography (CT) images,  possible morphologic changes occurring between CT and treatment,  patient mis-positioning and  organ motion during treatment itself. All these effects give an overall uncertainty of the order of few millimeters that can be larger than the dimension of the peak of the dose release (Bragg Peak). A precise monitoring of the dose is then essential for a good quality control of the treatment. Furthermore, the dose monitoring  would be particularly useful if provided during the treatment (in-beam monitoring) in order to provide a fast feedback to the beam. \\
Several methods have been developed to determine the Bragg Peak position online by exploiting the secondary particle production induced by the hadron beam. Since the irradiation of tissues with hadron beams produces nuclear excitations followed by photon emissions from de-excitation within few $\nano\second$ (prompt photons), both rate and production region could be used to monitor the dose~\cite{Testa}.\\
 
In this paper we present the measurement of secondary particles produced by the Poly-methyl methacrylate (PMMA) during an irradiation with carbon ions. In section~\ref{Catania}, we describe the setup of the experiment performed at the Laboratori Nazionali del Sud (LNS) Istituto Nazionale di Fisica Nucleare (INFN) di Catania with cerium-doped lutetium-yttirum ortho-silicate (LYSO) crystals for the detection of prompt photons emitted in the interaction of a $80\ \mega\electronvolt/$u fully stripped carbon beam with a PMMA target. The obtained results on the measured spectra and rates are then reported in sections~\ref{Risultati} and~\ref{Rates}.

\section{Experimental setup}
\label{Catania}

\begin{figure}[!ht]
\begin{center}
\includegraphics [width = 1 \textwidth] {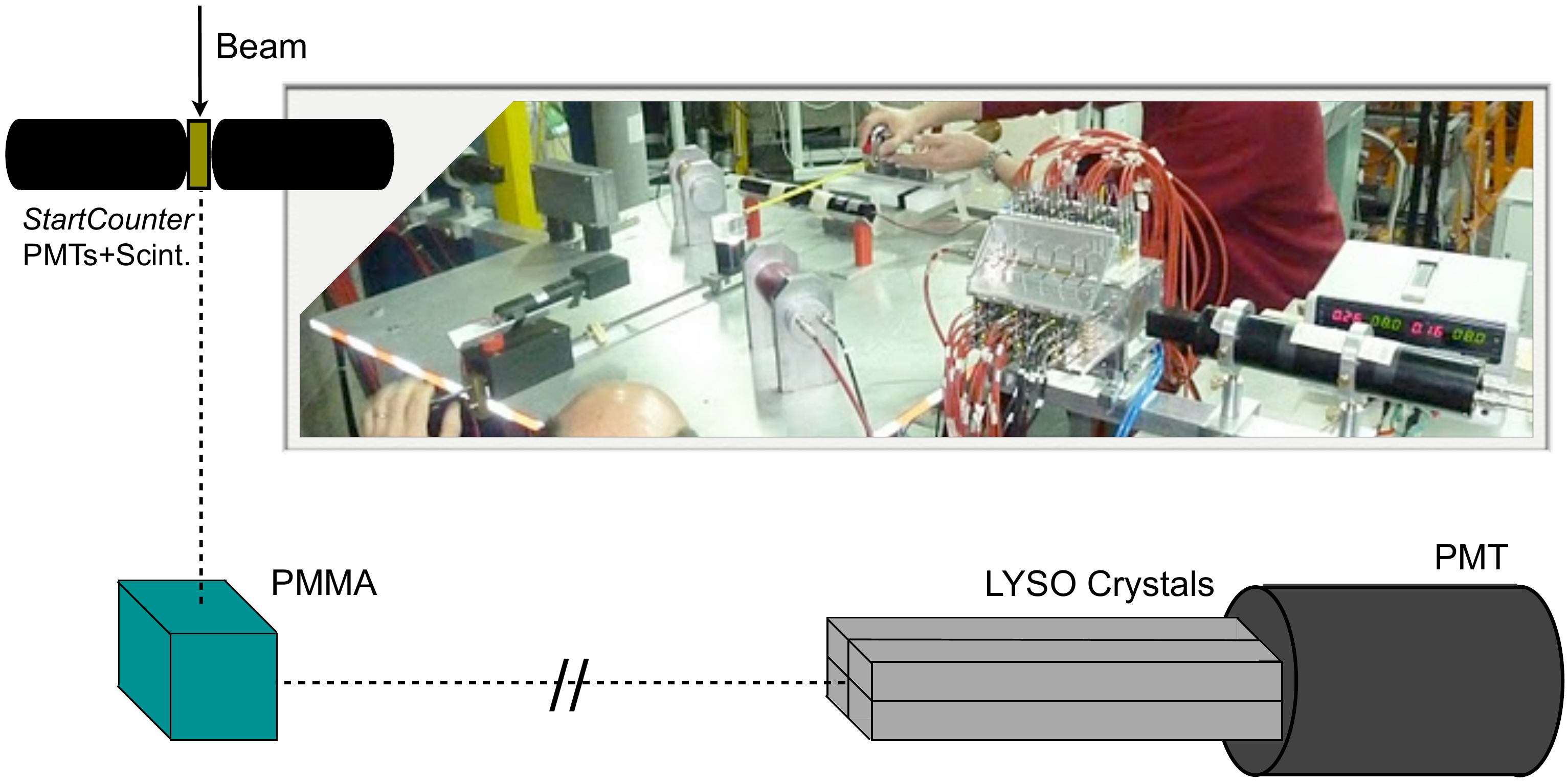}
\caption{\small{Schematic view of the experimental setup; the acquisition is triggered by the Start Counter in coincidence with the LYSO. In the picture of the detectors is also shown.}}
\label{fig:Schema}
\end{center}
\end{figure}

The experimental setup is shown in Fig.~\ref{fig:Schema}. A $4\times 4 \times 4\ \centi\meter^3$ PMMA target is placed on $960\ \mega\electronvolt$ ($80\ \mega\electronvolt/$u) fully stripped $^{12}C$ ion beam. The beam rate, ranging from hundred $\kilo\hertz$ to $\sim 2 \mega\hertz$, is monitored with a $1.1\ \milli\meter$ thin scintillator on the beam line read with two photomultiplier tubes (PMT) (Hamamatsu $10583$) put in coincidence (Start Counter) and placed at  $17\ \centi\meter$ from the PMMA. 

Prompt photons are discriminated by measuring a time difference between the beam impact and the photon detection consistent with a particle emitted istantaneously and traveling at the speed of light.
The main background to prompt photons is due to neutrons, that are also emitted by nuclear interactions of the carbon beam in the PMMA. Since such neutrons are non relativistic, their arrival time is a good discriminator between photons and neutrons, hence the time resolution becomes critical in the signal selection. Therefore, we have chosen the LYSO crystals and tested their performance in discriminating between prompt photons and background neutrons.
LYSO crystals have high light output, high density, and fast decay time (as an indication see Tab.~\ref{tab:Cristalli}) and are commonly exploited in PET imaging, although in smaller sizes.

\begin{table}[ht]
\begin{center}
\begin{tabular}{l|c}
\hline 
\bf{LYSO Characteristics} & \\
\hline 
Effective Atomic Number  & $66$ \\
Density ($g/\centi\meter^3$) & $7.4$ \\
Radiation Length ($\centi\meter$) & $1.10$\\
Decay Constant ($\nano\second$)  & $40-44$ \\
Peak Emission ($\nano\meter$) & $428$ \\
Light Yield $\%$ NaI (Tl) & $75$\\
Index of Refraction & $1.82$\\
\hline
\end{tabular}
\end{center}
\caption{\small{Properties of LYSO crystals.}} 
\label{tab:Cristalli}
\end{table}

An array of $4$ LYSO crystals, each measuring $1.5 \times 1.5 \times 12\ \centi\meter^3$, is placed at $90\degree$ with respect to the beam line, at $74\ \centi\meter$ from the PMMA center. The scintillation light of the crystals is detected with a PMT (EMI $9814$B) triggered in coincidence with the Start Counter. A $12$-bit QDC (Caen $792$N) and a $19$-bit TDC (Caen $1190$B) provide the measurements of both the particles' energy and arrival time.

\subsection{Calibration and Simulation}
\label{Calib}

In order to calibrate the LYSO detector we used $^{22}Na$ and $^{60}Co$ sources: fitting gaussian peaks for the observed energy spectra, linearity is verified up to the energy of the second $^{60}Co$ peak ($1.33\ \mega\electronvolt$). At higher energies, when the pair-production mechanism becomes relevant in the shower development inside the crystals, there is a non-negligible probability to escape the detector for one or both  $511\ \kilo\electronvolt$ photons produced by the positron annihilation, mechanism known as single- and double- escape respectively. When such phenomena occur, the output signal appears as a superposition of three gaussian structures and, therefore, the measured energy is biased and its resolution degrades.

This effect was simulated using the GATE framework~\cite{Gate}, a tool dedicated to medical imaging, radiotherapy and hadrontherapy, based on the GEANT4 Monte Carlo (MC) code~\cite{Geant1},\cite{Geant2}.

\begin{figure}[!ht]
\begin{center}
\includegraphics [width =1 \textwidth] {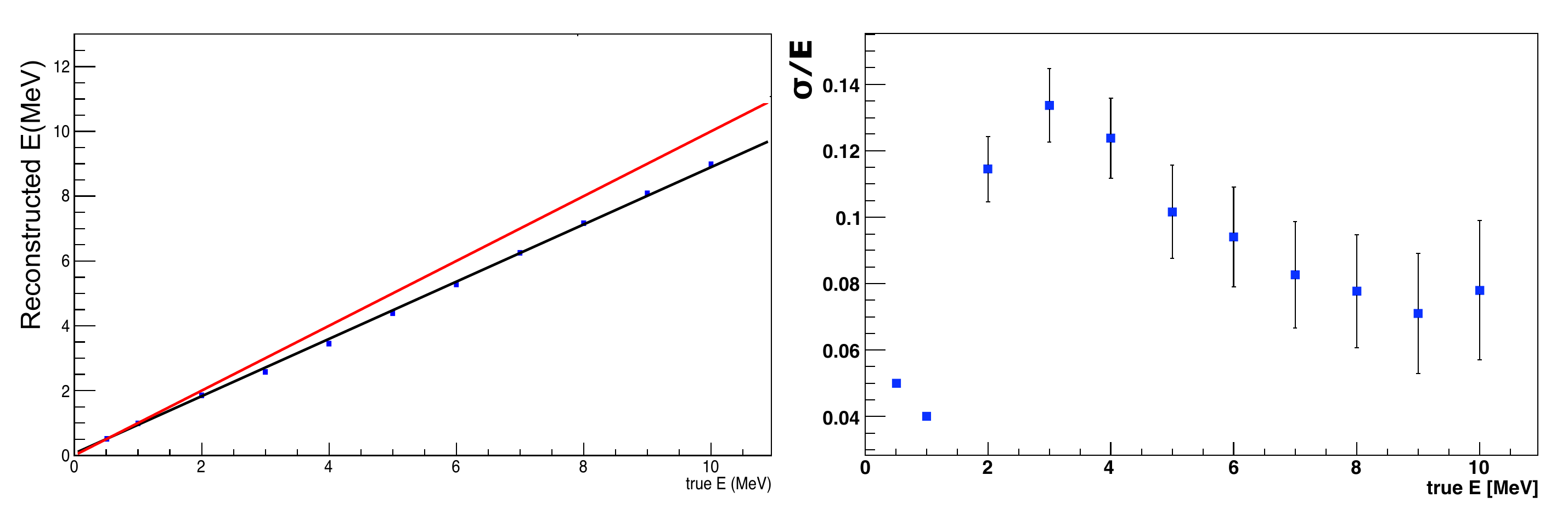}
\caption{\small{Results obtained using GATE simulation. Left (right): mean values (resolution/energy) of the energy spectra recorded in LYSO detector, as a function of the photon energy. The red line in the left picture represents the bisector line.}}
\label{fig:Calib2}
\end{center}
\end{figure}

We simulated the response of the experimental setup to monochromatic photons of several energies. The simulated distribution of the collected photoelectrons have been fitted with gaussian functions with exponential tails, and the mean values for energies below $1\ \mega\electronvolt$ were used for energy calibration, thus following the same procedure used for data analysis. Using simulated photons of energies above $1\ \mega\electronvolt$ we estimate the bias on the energy measurement and the degradation of the resolution due to single- and double- escapes. The resolution obtained for the calibration peaks determines the number of photons to be used in the MC simulation. The results are shown in Fig.~\ref{fig:Calib2}. The left plot shows the mean value of the energy reconstructed simulating the response of the experimental setup as a function of the true energy; the comparison with the bisector (red line) shows the $\sim 10\%$ bias on the energy measurement  due to single- and double- escapes for $E>1\ \mega\electronvolt$. On the right plot, the same results are shown for the $\sigma / E$ parameter as a function of the true energy. A deterioration of the resolution for $E>1\ \mega\electronvolt$ is observed.

\subsection{Event timing}
\label{NC}

In order to select signal events, the time difference between the energy deposition in the LYSO  detector ($T_{LYSO}$) and arrival time of the $^{12}C$ ions on the Start Counter ($T_{SC}$), referred to as $\Delta t$, is evaluated.  When both photo-multipliers of the Start Counter yield a signal,  the smallest absolute value of $\Delta t$ is chosen. 

The number of carbon ions reaching the target and the electronics inefficiency due to the acquisition  \textit{dead-time} are needed to estimate the prompt photon rate.
The number of carbon ions reaching the PMMA in a given time interval is computed by counting the number of signals given by the Start Counter ($N_{SC}$) within randomly-triggered time-windows of $T_w=2\ \micro\second$. Measuring the total run time ($T_{tot}$) and the number of time windows considered ($N_w$), the number of carbon ions is estimated as 

\begin{equation}
N_C=\frac{N_{SC}}{\epsilon_{SC}}\frac{T_{tot}}{N_w*T_w}.
\end{equation}

The Start Counter efficiency $\epsilon_{SC}=(96\pm 1)\%$ has been estimated by exploiting the  two-sided PMT readout with negligible dark counts: $\epsilon_{SC}= \epsilon_{sc1} \cdot \epsilon_{sc2} = N_{12}^2/(N_1 \cdot N_2)$, with $N_{1,2}$ are the single PMT counts and $N_{12}$ their coincidences.

To correct $N_C$ for the dead-time inefficiency, $\epsilon_{DT}$ has been estimated from the total acquisition time ($T_{dead}$) as:

\begin{equation}
\epsilon_{DT}=1-\frac{T_{dead}}{T_{tot}}.
\end{equation}

The measured values of $\epsilon_{DT}$ at an average carbon ion rate $R_C=N_C/T_{tot}$ ranges from  $70\%$ at $0.6\ \mega\hertz$ up to $47\%$ at $2\ \mega\hertz$. This efficiency correction was then applied to data.

\section{Prompt photon energy spectrum measurements}
\label{Risultati}

\begin{figure}[!ht]
\begin{center}
\includegraphics [width =1 \textwidth] {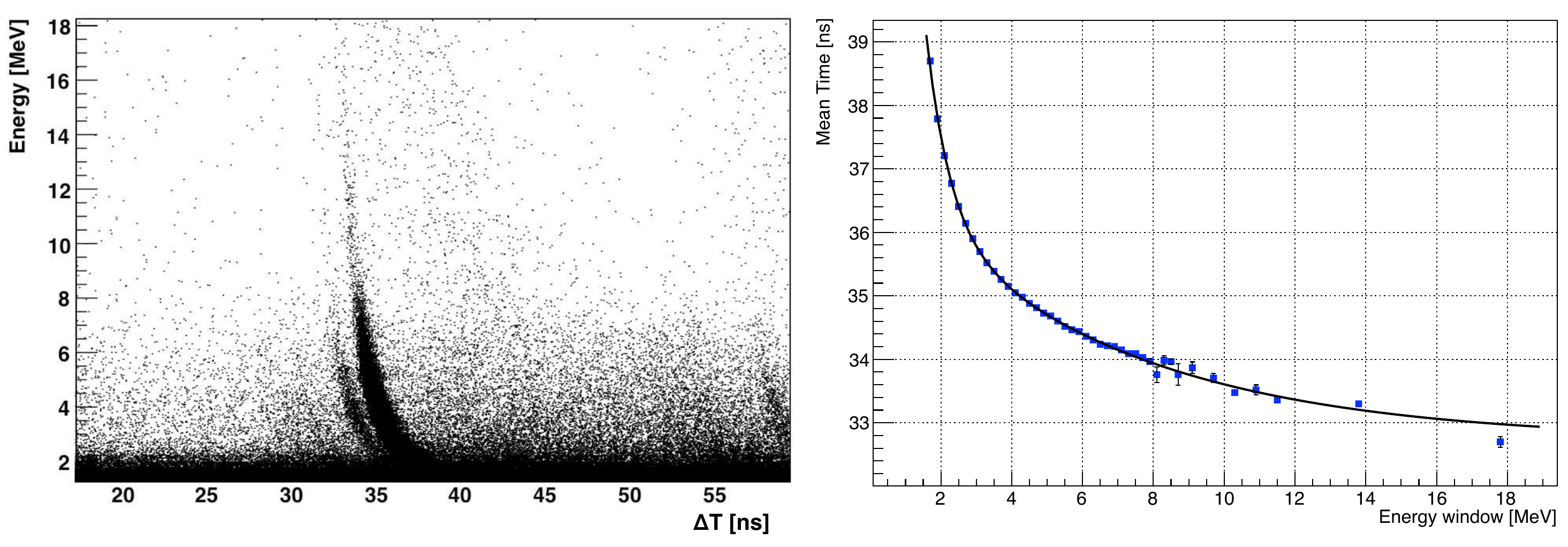} 
\caption{\small{Left: Calibrated energy released in LYSO crystal as a function of the arrival time, $\Delta t$. Three component are present: an horizontal band due to the LYSO intrinsic noise; an almost vertical band due to the signal from the prompt photons; a diffused cloud mainly due to neutrons at $\Delta t$ values larger than those of the prompt photons population. Right: estimated time slewing correction.}}
\label{fig:Banana}
\end{center}
\end{figure}

The correlation between the reconstructed energy (E) and the measured $\Delta t$ is shown in Fig.~\ref{fig:Banana} (Left). Three populations are present: a horizontal band from LYSO intrinsic noise; an almost vertical band due to the signal from the prompt photons; a diffused cloud mainly due to neutrons at $\Delta t$ values larger than those of the prompt photons population. Events are shown in a $\pm 23$ \nano\second\ window around the average $\Delta t$ values for prompt photons to avoid events associated with the previous and subsequent cyclotron cycles. 
The measured horizontal offset in the $\Delta t$ distribution for prompt photons is due to the electronic setup. A secondary component, few $\nano\second$ before the most populated band, is visible: this distribution is due to the prompt photons produced inside the Start Counter. The earlier arrival time of these events is due to their shorter travel.
The shape of the prompt photon band is instead due to the time slewing effect induced by the  front-end electronics fixed voltage threshold.
This effect can be corrected for by fitting the $\Delta t$ distribution in bins of $E$, as shown in Fig.~\ref{fig:Banana} (right), thus obtaining the correction $C(E)$. 

\begin{figure}[!ht]
\begin{center}
\includegraphics [width =1 \textwidth] {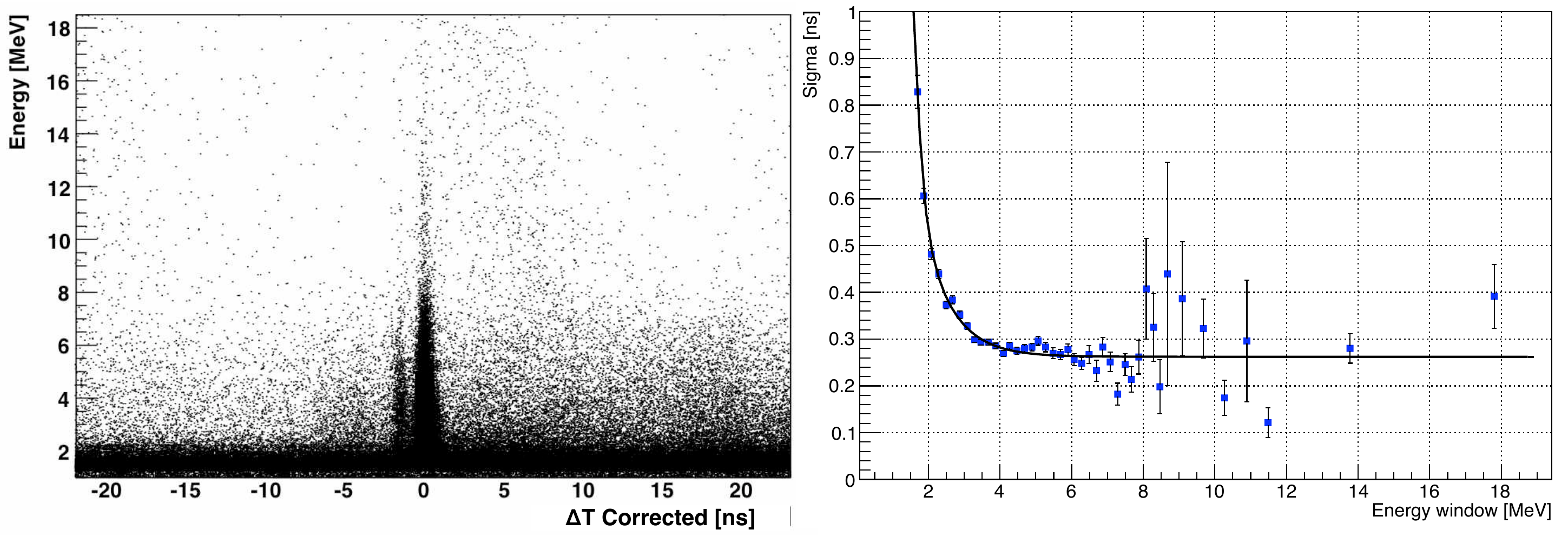} 
\caption{\small{Left: calibrated energy released in LYSO crystal as a function of the arrival time corrected taking into account the slewing effect. Right: measured time resolution as a function of the calibrated energy.}}
\label{fig:BananaCorr}
\end{center}
\end{figure}

Fig.~\ref{fig:BananaCorr} (Left) shows the energy spectrum as a function of the corrected time ($\Delta t_{corr}=\Delta t-C(E)$).
Fig.~\ref{fig:BananaCorr} right shows the time resolution ($\sigma_{\Delta t}$) as a function of the energy. A resolution better than $300\ \pico\second$ is achieved for energies greater than $3\ \mega\electronvolt$.

\begin{figure}[!ht]
\begin{center}
\includegraphics [width =0.5 \textwidth] {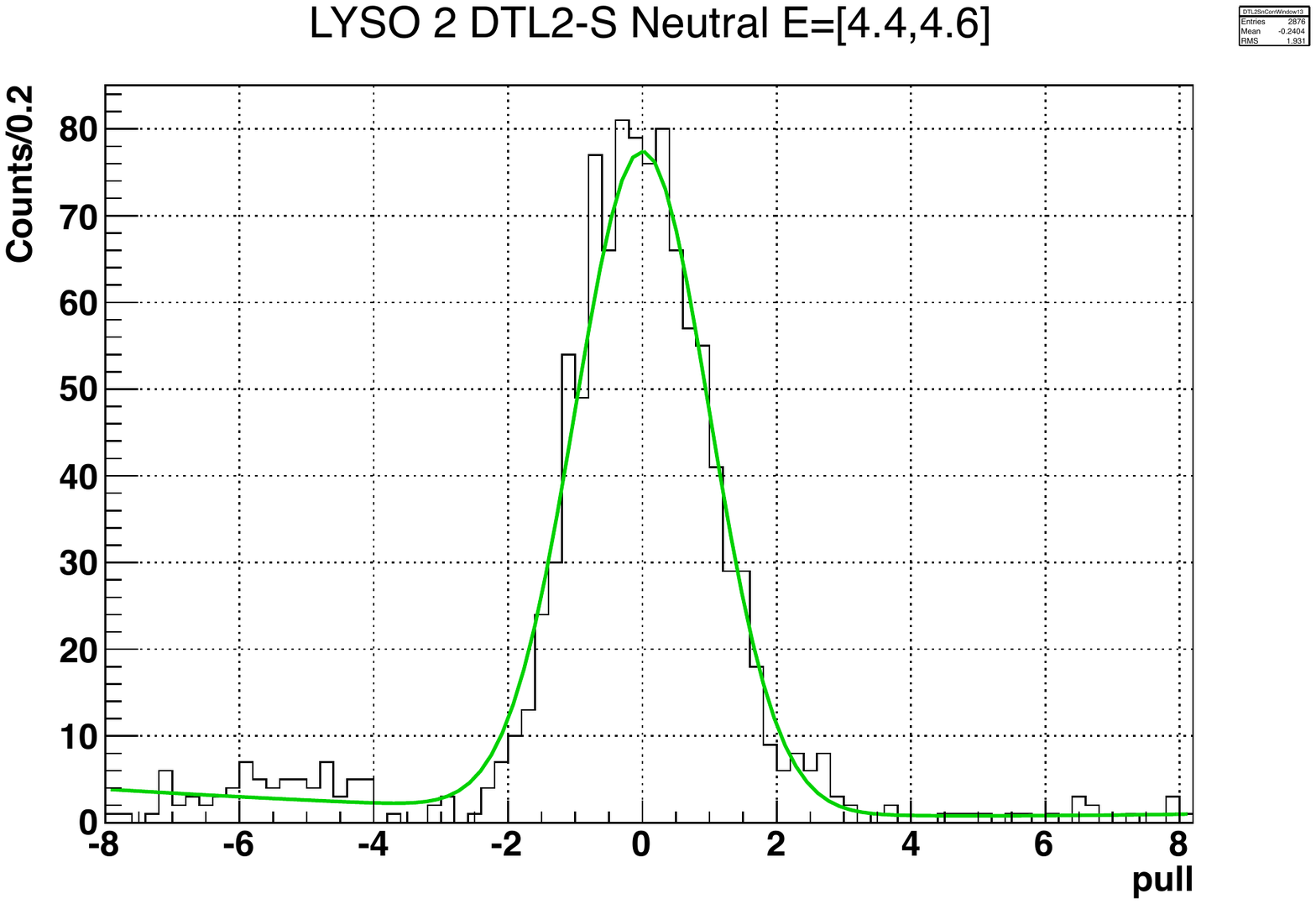}
\caption{\small{Example of a pull distribution and the corresponding fit to extract the number of prompt photons.}}
\label{fig:Pull}
\end{center}
\end{figure}

\begin{figure}[!ht]
\begin{center}
\includegraphics [width =0.51 \textwidth] {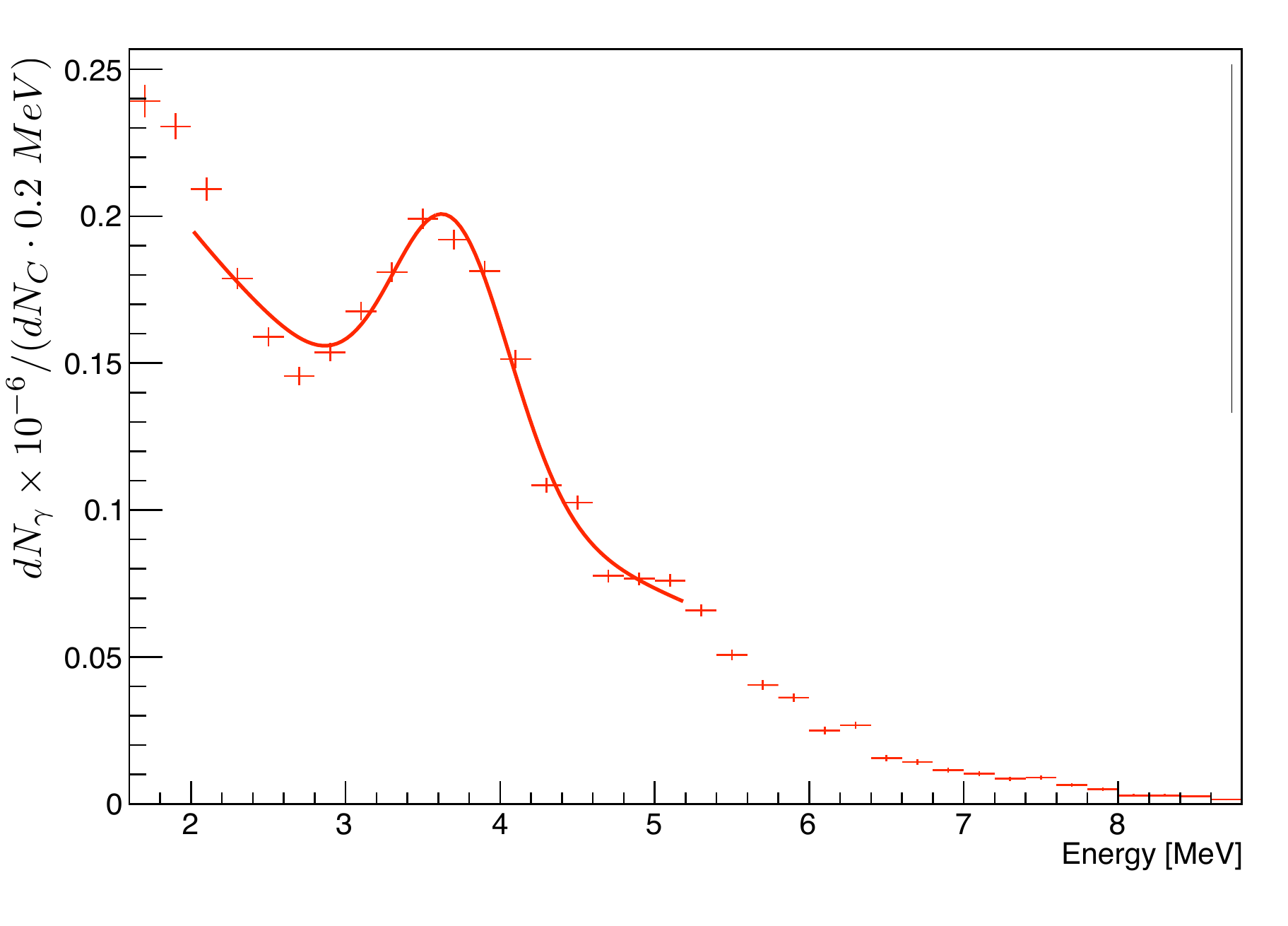}
\includegraphics [width =0.48 \textwidth] {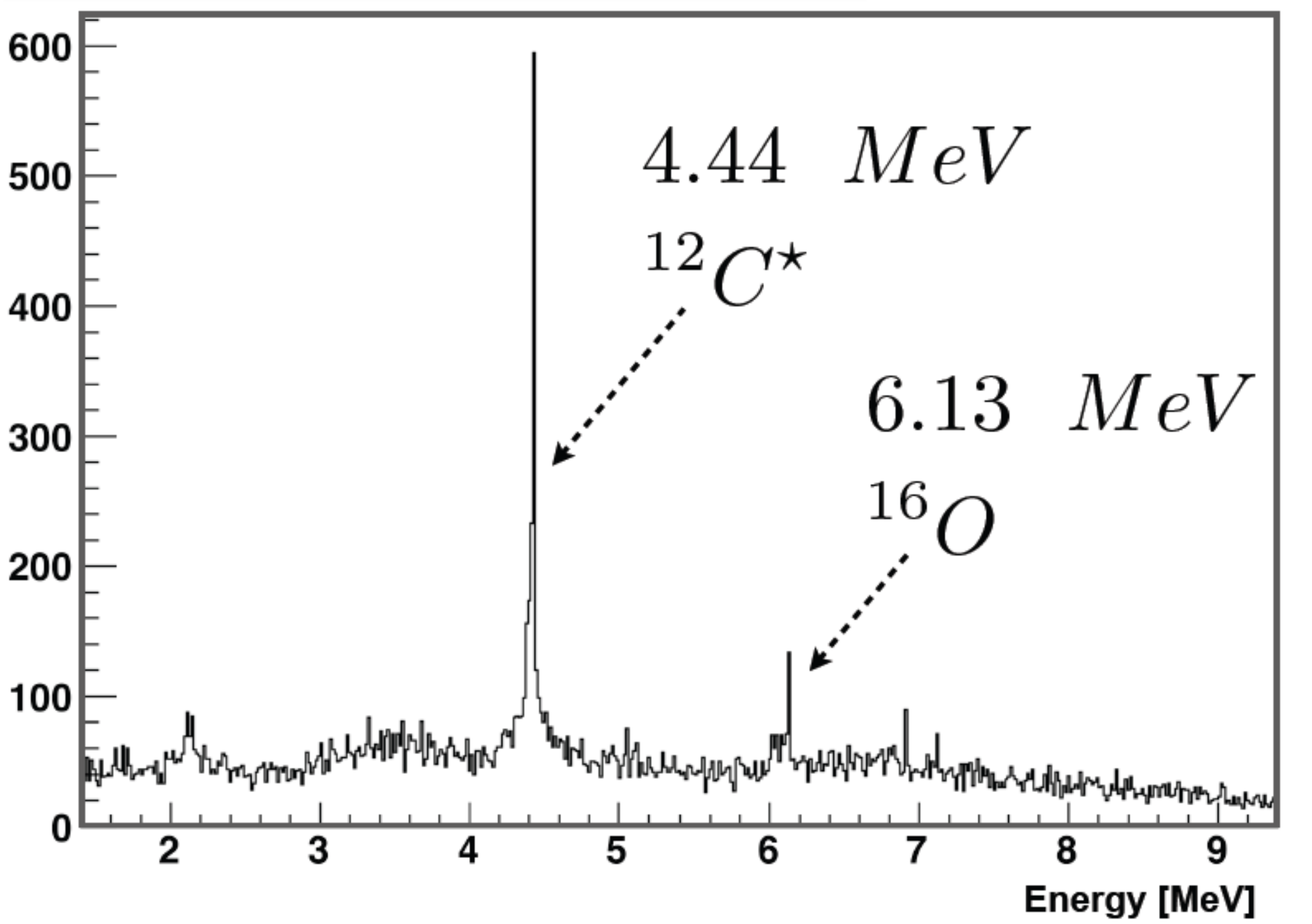}
\caption{\small{Left: Measured prompt photons spectra obtained with the LYSO detector. The fit of the  peak gives a fraction of $f_{^{12}C}=(13.9\pm 0.6)\%$ for prompt photons over carbon ions at $4.44\ \mega\electronvolt$. Right: Monte Carlo energy spectrum of the prompt photons in the detector acceptance without detector response simulation.}}
\label{fig:Spec}
\end{center}
\end{figure}

In order to measure the prompt photon energy spectrum originating from the PMMA, the number of prompt photons in each energy bin is extracted from the fit to the distribution of $\Delta t_{corr}/\sigma_{\Delta t_{corr}}$ (pull) in that bin: the fit uses a polinomial function and a gaussian distribution, fixing the width at one and the mean at zero. Fig.~\ref{fig:Pull} shows an example of the pull distribution obtained for one energy bin. The area under the gaussian provides the number of prompt photons in a given energy bin.
The resulting measured spectrum, normalized to the number of incident carbon ions and corrected for $\epsilon_{DT}$ is shown in Fig.~\ref{fig:Spec} (left). 

The experiment setup has been simulated with GATE to obtain the expected energy spectrum :  we used G$4$QMDReaction to model ions$'$ inelastic processes and the interactions between $^{12}C$ and the PMMA. The simulated spectrum of the prompt photon emitted in the geometrical acceptance of the LYSO crystal is shown in Fig.~\ref{fig:Spec} (Right).
Without the detector effects, two components are visible: a continuum component and few excitation peaks. The two most prominent structures at $E=4.44\ \mega\electronvolt$ and $6.13\ \mega\electronvolt$ come from $^{12}C^*$ and $^{16}O^*$ excited states decays respectively.

Fitting the data using exponential function for the continuum contribution and a gaussian distribution for the main $^{12}C^*$ peak we obtain for this peak a mean value and a resolution compatible within errors with the expectations from the simulation:  $ \sigma_{E} \sim 0.5\ \mega\electronvolt$,  for $E=4.44\ \mega\electronvolt$  (see Fig.~\ref{fig:Calib2}). From this fit we measure a contribution from the $^{12}C^*$ decays line  to the overall rate of prompt photons with energies $E>2\mega\electronvolt$ equal to $f_{^{12}C}=(13.9\pm 0.6)\%$.

\begin{figure}[!ht]
\begin{center}
\includegraphics [width =0.6 \textwidth] {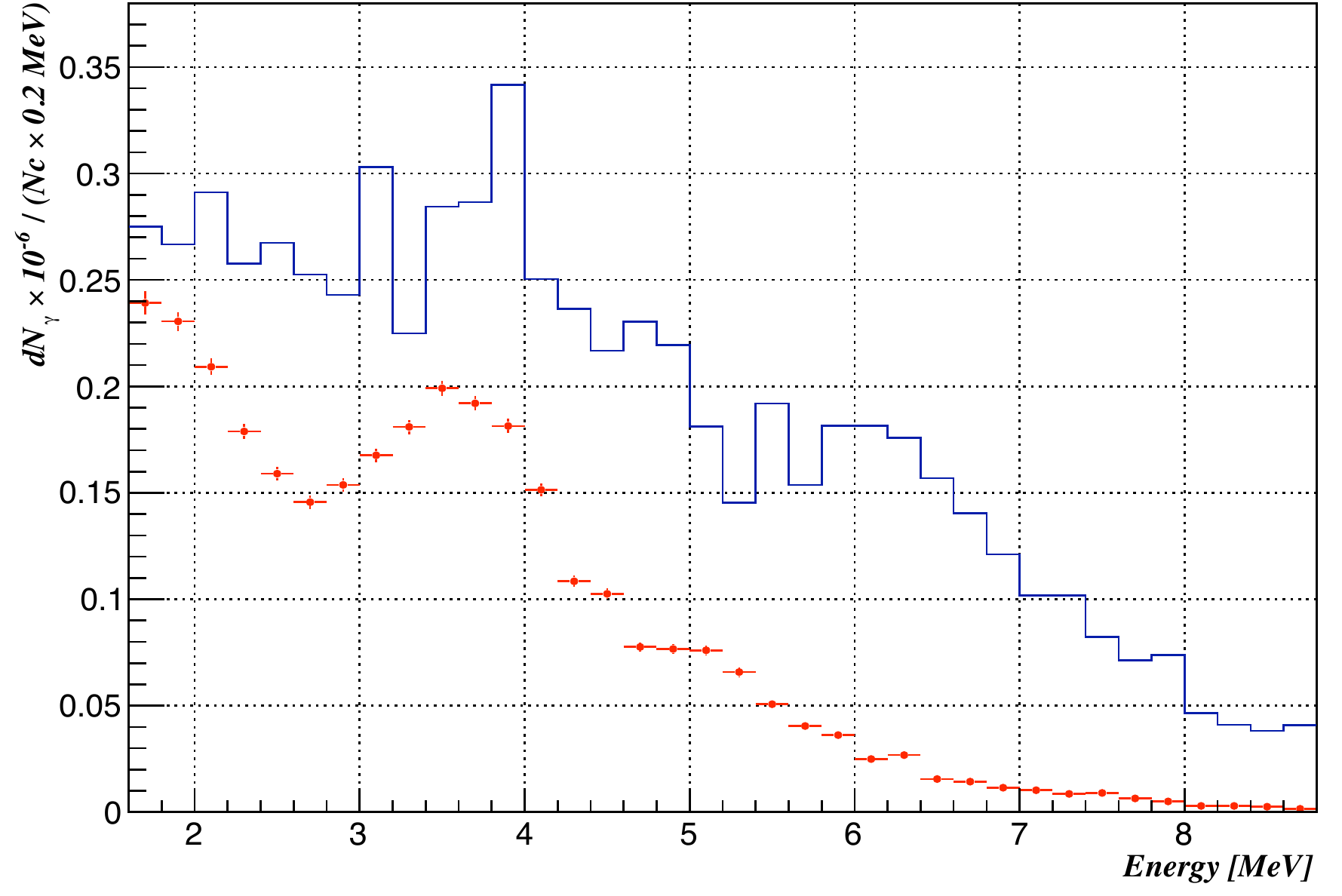}
\caption{\small{Data-MC comparison of the observed energy spectrum of the prompt photons. All spectra are normalized to the number of incident carbon ions.}}
\label{fig:Sovrapposizione}
\end{center}
\end{figure}

The MC true energy distribution has been folded with the detector response (detector and Start Counter efficiency, resolution, and bias), using the simulations described in Sec.~\ref{Catania}. The folded MC spectrum is compared with the measured one in Fig.~\ref{fig:Sovrapposizione}, with the spectra normalized to the number of incident carbons. The comparison shows the level of agreement between data and MonteCarlo simulation: both the normalization and the fraction of $E=4.44\ \mega\electronvolt$ photons require further investigation.

\section{Rate measurements}
\label{Rates}

\begin{figure}[!ht]
\begin{center}
\includegraphics [width =0.8 \textwidth] {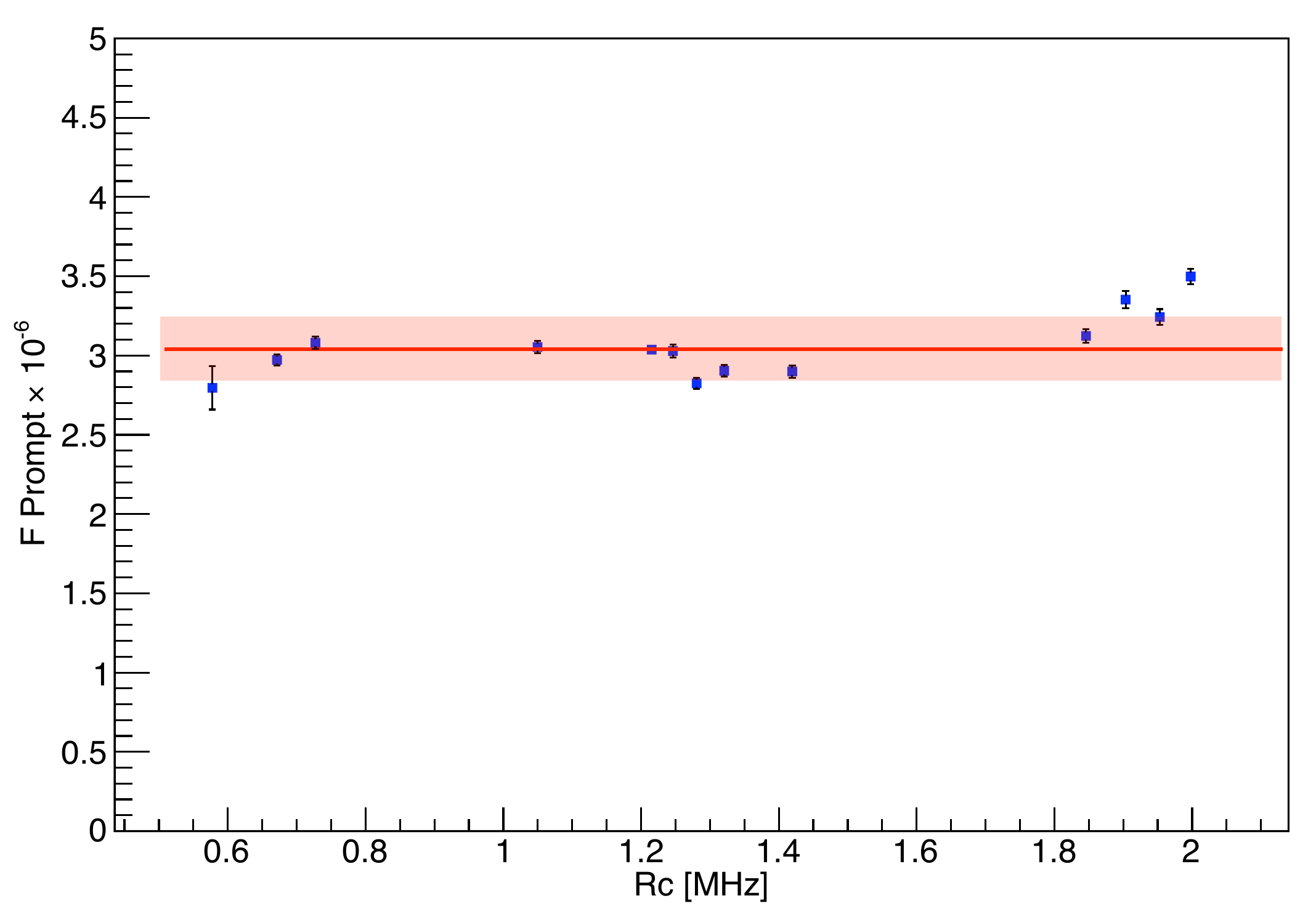}
\caption{\small{Rate of prompt photons with $E>2\ \mega\electronvolt$ normalized to the rate of carbon ions as a function of the latter. The red fit line is shown with the error band (both statistical and systematic errors are taken).}}
\label{fig:dosemeas}
\end{center}
\end{figure}

The ultimate goal of the studies on prompt photons is the measurement of the number of carbon ions delivered to the patient.
To understand the potentiality of LYSO detectors as dosimeters for prompt photons measurements, we select only events with $E>2\ \mega\electronvolt$ rejecting the background from natural radioactivity of the LYSo crystal. For such events we proceed with a side-band subtraction on the prompt photon $\Delta t_{corr}$ distribution and extract the number of prompt photons, $N_{meas}$. For runs with different carbon ion rates we then measure the ratio between the rates of observed prompt photons ($R_{prompt}$) and carbon ions ($R_c$):
 
 \begin{equation}
F_{prompt}=\frac{R_{prompt}}{R_C}=\frac{N_{meas}}{\epsilon_{DT}\epsilon_{SC}N_C}
\end{equation}

with $\epsilon_{DT}$,  $\epsilon_{SC}$, and $N_C$ the dead-time efficiency, the Start Counter efficiency, and the number of carbon ions respectively, measured on data (Sec.~\ref{Catania}). 
Fig.~\ref{fig:dosemeas} shows that $F_{prompt}$ does not dependent on the carbon rate and averages to

\begin{equation}
F_{prompt}=(3.04 \pm 0.01_{stat}\pm0.20_{sys})\times 10^{-6}. 
\end{equation}

The systematic error was introduced to account for the dispersion of  the values which is well above the statistical fluctuations. Under the hypothesis of a flat distribution, it was estimated as the semi-dipersion divided by $\sqrt{3}$.

We then measured the differential prodution rate:

\begin{equation}
\frac{dN_{\gamma}}{dN_Cd\Omega}(E>2\mega\electronvolt, \theta=90^o)=\frac{F_{prompt}}{\epsilon_{meas} \Omega_{LYSO}}
\end{equation}

with $\Omega_{LYSO}=1.3\times 10^{-4}$ the solid angle covered by the detector and $\epsilon_{meas}=(81.3\pm2.5)\%$ the detection efficiency estimated with the 
Monte Carlo simulation (Sec.~\ref{Catania}). To evaluate the uncertainty due to the assumed energy distribution of the photons, this efficiency has been computed both using a flat energy distribution and using the simulated MC spectrum. The semi-difference between the two results is then assumed as the uncertainty.
 
The final result for the differential rate of prompt photons is: 
\begin{equation}
\frac{dN_{\gamma}}{dN_Cd\Omega}(E>2\mega\electronvolt, \theta=90^o)=(2.92\pm 0.01_{stat} \pm 0.19_{sys})\times 10^{-2} sr^{-1}.
\end{equation}

\section*{Conclusions} 

The measurement of the energy spectrum and rate of prompt photons produced in the  interaction of $80\ \mega\electronvolt/$u carbon ions with a PMMA target has been reported. These results were obtained with a LYSO detector with a time resolution of $300\ \pico\second$ for photons with energies above $3\ \mega\electronvolt$. The excellent time resolution was used to discriminate the neutrons and to determine a precise energy spectrum of the prompt photons. In particular, we measure the fraction of prompt photons with energies $E>2\ \mega\electronvolt$ coming from the $E=4.44\ \mega\electronvolt$ line to be $f_{^{12}C}=(13.9\pm 0.6)\%$.

We also measure the rate of photons per carbon ions triggered to be $F_{prompt}=(3.04 \pm 0.01_{stat}\pm0.20_{sys})\times 10^{-6} $. We also proved $F_{prompt}$ is  largely independent, after dead-time corrections, from the carbon rate in the range between $0.6$ and $2\ \mega\hertz$ explored in this experiment.

Finally, we measured the differential production rate to be $dN_{\gamma}/dN_Cd\Omega \ \ (E>2\ \mega\electronvolt$, $\theta=90\ \degree)=(2.92\pm 0.19)\times 10^{-2}$ sr$^{-1}$.

\begin{center}\textbf{\large Acknowledgements}\end{center}{\large \par}

The authors would like to thank Dr. M.~Pillon and Dr. M.~Angelone (ENEA-Frascati, Italy) for allowing us to validate the response of our detector to neutrons on the Frascati Neutron Generator; C.~Piscitelli (INFN-Roma, Italy) for the realization of the mechanical support. The  staff of the INFN-LNS (Catania, Italy) test beam is gratefully acknowledged for their kind cooperation and helpfulness.

\end{document}